\documentclass[acmsmall,nonacm]{acmart}

\acmJournal{PACMPL}
\acmVolume{1}
\acmNumber{CONF} 
\acmArticle{1}
\acmYear{2018}
\acmMonth{1}
\acmDOI{} 
\startPage{1}

\setcopyright{none}

\usepackage{booktabs}   
\usepackage{subcaption} 

\usepackage{xcolor}

\usepackage{prettyref}
\newcommand{\pref}{\prettyref}
\newrefformat{thm}{Theorem~\ref{#1}}
\newrefformat{cor}{Corollary~\ref{#1}}
\newrefformat{lem}{Lemma~\ref{#1}}
\newrefformat{cha}{Chapter~\ref{#1}}
\newrefformat{sec}{Section~\ref{#1}}
\newrefformat{app}{Appendix~\ref{#1}}
\newrefformat{tab}{Table~\ref{#1}}
\newrefformat{fig}{Figure~\ref{#1}}
\newrefformat{alg}{Algorithm~\ref{#1}}
\newrefformat{exa}{Example~\ref{#1}}
\newrefformat{def}{Definition~\ref{#1}}
\newrefformat{li}{Line~\ref{#1}}
\newrefformat{eq}{Equation~\ref{#1}}

\usepackage{listings}

\usepackage[prefix=tikzsym]{tikzsymbols}
\usepackage{svg}
\newcommand\emoji[1]{\includesvg[height=10pt]{emojis/#1.svg}}

\begin{document}

\title{System-Specific Interpreters Make Megasystems Friendlier}

\author{Matthew Sotoudeh}
\orcid{0000-0003-2060-1009}             
\affiliation{
  \institution{Stanford University}            
  \city{Stanford}
  \state{CA}
  \country{USA}                    
}
\email{sotoudeh@stanford.edu}

\begin{abstract}
    Modern operating systems, browsers, and office suites have become
    \emph{megasystems} built on millions of lines of code. Their sheer size can
    intimidate even experienced users and programmers away from attempting to
    understand and modify the software running on their machines.
    This paper introduces \emph{system-specific interpreters} (SSIs) as a tool
    to help users regain knowledge of and control over megasystems.
    SSIs directly execute individual modules of a megasystem in a
    \texttt{gdb}-like environment without forcing the user to build, run, and
    trace the entire system.
    %
    %
    %
    A prototype framework to help write SSIs is described in this paper and
    available for download at~\url{https://github.com/matthewsot/ssi-live22}.
\end{abstract}

\maketitle

\section{Introduction}
Ballooning user expectations have caused computer systems to grow in size and
complexity over time.  Operating systems, web browsers, office suites,
compilers, and even websites are routinely hundreds of thousands or even
millions of lines of code across a variety of programming languages.

These \emph{megasystems} are extremely useful, surprisingly dependable, and
often free software. But their sheer size makes it increasingly difficult for
users and hobbyist programmers to understand and modify them. Even
\emph{compiling} some megasystems, such as the Linux kernel, can be a
surprisingly challenging task. Such barriers represent an unfortunate
bottleneck through which even reasonably motivated and skilled users are
prevented from understanding and exercising control over the software that has
become necessary for their daily lives.

This paper proposes the development and use of \emph{system-specific
interpreters} (SSIs) to help users understand and modify portions of such
megasystems.  The key insight is that, from the perspective of any given module
of the larger system, the rest of the system exposes a \emph{domain-specific
language} in which the module is written. Directly building an interpreter (the
SSI) for that language allows the user to run individual modules, files, and
functions without having to wait for the system to build, construct inputs
triggering the code in question, or insert tracing code.

In fact, programmers already build such SSIs \emph{informally} in their heads
when attempting to understand megasystems. We usually start at some interesting
part of the code, e.g., a particular device driver, and read it \emph{without}
knowing exactly how it is called by the overall megasystem or what all of the
functions it calls do. Over time, an informal mental model of this interface
between the module and megasystem is formed to enable such underconstrained
reasoning.

Our key idea is to \emph{formalize and mechanize} this mental model of the
interface between the module and the rest of the system. By giving users the
tools to formally specify (as an SSI) their previously informal model of the
system--module interface, the cognitive burden of remembering a large model is
reduced. Executing code against this model via the SSI helps users find
inconsistencies or inaccuracies in their mental model of the system.

We describe a framework that makes writing SSIs easier. The key idea is to
separate implementation of \emph{language-level} syntax and semantics from
\emph{system-level} syntax and semantics. Once written, SSIs simulate
execution of a module \emph{in isolation}, similar to a unit test.  The user
interacts with the SSI's \texttt{gdb}-style interface to trace and modify
intermediate states of the execution. After modifying the code, the SSI can be
re-run to check if the modification had the desired effect. By directly
interpreting the specific code that the programmer cares about, SSIs are a
first step towards bringing the power of live programming and REPL environments
to bear on the unique challenges of lower-level systems code.

\section{Case Study}
Consider a curious Jane Doe. One day, she is using a Raspberry Pi to control a
light strip and becomes interested in how the kernel's GPIO driver works. She
pulls up the Linux source code, and finds the relevant \texttt{pinctrl-bcm2835}
driver code excerpted in~\pref{fig:bcm2835-code}. Reading the code, she notices
that the broader Linux kernel provides essentially a domain-specific language
(DSL) to the driver code. This DSL includes, for example, commands to remap
(\texttt{devm\_ioremap\_resource}) or write to (\texttt{writel}) memory-mapped
input-output (MMIO) addresses. The DSL exposed by the kernel actually spans
multiple real languages, e.g., MMIO addresses are implicitly read from a device
tree source include (DTSI) file that gets compiled into the kernel at build
time.

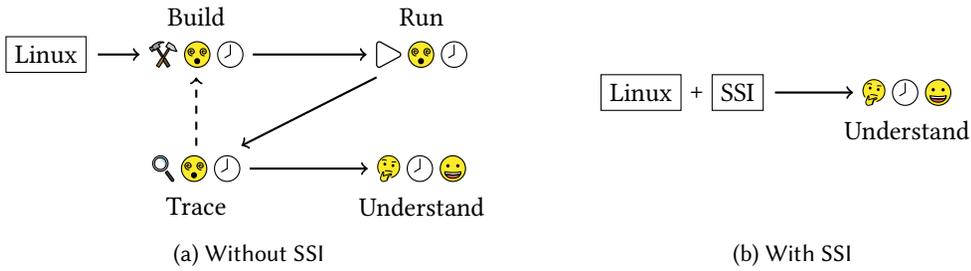
\begin{figure}
    \centering
    \begin{subfigure}[b]{0.48\linewidth}
        \centering
        \begin{tikzpicture}
            \node[rectangle,draw=black] (a) at (0, 0) {Linux};
            \node (b) at (2, 0) {\emoji{hammer-and-pick} \emoji{face-with-spiral-eyes} \emoji{eight-oclock}};
            \node at (2, 0.5) {Build};

            \node (c) at (5, 0) {\emoji{play-button} \emoji{face-with-spiral-eyes} \emoji{eight-oclock}};
            \node at (5, 0.5) {Run};

            \node (d) at (2, -1.5) {\emoji{magnifying-glass-tilted-left} \emoji{face-with-spiral-eyes} \emoji{eight-oclock}};
            \node at (2, -2) {Trace};

            \node (e) at (5, -1.5) {\emoji{thinking-face} \emoji{eight-oclock} \emoji{grinning-face}};
            \node at (5, -2) {Understand};

            \draw[thick,->] (a.east)++(1.5mm, 0) -- (b);
            \draw[thick,->] (b) -- (c);
            \draw[thick,->] (c) -- (d);
            \draw[thick,->] (d) -- (e);

            \draw[thick,->,dashed] (d) -- (b);
        \end{tikzpicture}
        \caption{Without SSI}
    \end{subfigure}
    \hfill
    \begin{subfigure}[b]{0.48\linewidth}
        \centering
        \begin{tikzpicture}
            \node[rectangle,draw=black,anchor=east] at (0, 0) {Linux};
            \node at (0.2, 0) {+};
            \node[rectangle,draw=black,anchor=west] (a) at (0.4, 0) {SSI};

            \node (b) at (3, 0) {\emoji{thinking-face} \emoji{eight-oclock} \emoji{grinning-face}};
            \node at (3, -0.5) {Understand};

            \draw[thick,->] (a.east)++(1.5mm, 0) -- (b);
        \end{tikzpicture}
        \vspace{10mm}
        \caption{With SSI}
    \end{subfigure}
    \caption{Poking at megasystem code usually involves figuring out how to:
    (i) build the system, (ii) execute it with the right inputs to reach the
    desired code, and (iii) trace the effects of the code. None of these steps
    are easy in most modern megasystems. SSIs execute individual modules of a
    megasystem within an interpreter environment, allowing users to skip
    directly to the key poking-and-understanding phase of the cycle.}
    \label{fig:flowchart}
\end{figure}

Jane quickly recalls why playing with megasystem codebases is a daunting task.
Traditional debugging tools like \texttt{gdb} require her to build and run the
code, but Jane has only ever used a pre-built kernel. Even if she could build
the kernel or driver, Jane may not always have immediate access to a Raspberry
Pi to run the code on. Even if all of those issues were worked out, Jane would
have to spend time writing a test program that correctly invokes the right
combination of syscalls to trigger the driver code she is trying to understand,
and then somehow figure out what the driver code is actually doing \emph{to the
hardware}, i.e., what MMIO addresses are actually being written to at each
point in the program and with what values.  Unless Jane wants to try and
understand the code entirely in her head, without the benefit of \emph{actually
running it like code to see what it does}, she must solve all of these problems
before even beginning to understand the code.

\begin{figure}[t]
    \centering
    \begin{subfigure}[b]{0.48\linewidth}
        {\small From \texttt{pinctrl-bcm2835.c}:}
        \begin{lstlisting}[linewidth=0.9\linewidth,basicstyle=\tiny]
/* ... */
void bcm2835_gpio_wr(struct bcm2835_pinctrl *pc, unsigned reg, u32 val) {
	writel(val, pc->base + reg);
}
/* ... */
static void bcm2835_gpio_irq_enable(
    struct irq_data *data) {
  /* ... */
  unsigned gpio = irqd_to_hwirq(data);
  unsigned offset = GPIO_REG_SHIFT(gpio);
  /* ... */
}
/* ... */
int bcm2835_pinctrl_probe(
    struct platform_device *pdev) {
 /* ... */
 err = of_address_to_resource(np, 0, &iomem);
 pc->base = devm_ioremap_resource(dev, &iomem);
 /* ... */
}
    \end{lstlisting}
        {\small From \texttt{bcm283x.dtsi}:}
        \begin{lstlisting}[linewidth=0.9\linewidth,basicstyle=\tiny]
/* ... */
    gpio: gpio@7e200000 {
            compatible = "brcm,bcm2835-gpio";
            reg = <0x7e200000 0xb4>;
/* ... */
    \end{lstlisting}
    \caption{Snippets from the \texttt{pinctrl-bcm2835} driver}
    \label{fig:bcm2835-code}
    \end{subfigure}
    \begin{subfigure}[b]{0.48\linewidth}
        \newcommand\lstprompt[1]{\color{white}{\textbf{#1}}}
        \begin{lstlisting}[backgroundcolor={\color{black!90}},basicstyle=\tiny\color{gray!30},morekeywords={ssi,>},numbers=none,mathescape=true]
Loaded driver:
    Description: Broadcom BCM2835/2711 pinctrl and GPIO driver
    Author(s): Chris Boot, Simon Arlott, Stephen Warren
    License: GPL
Choose device:
0 : brcm,bcm2835-gpio
1 : brcm,bcm2711-gpio
2 : brcm,bcm7211-gpio
Choice: 0
$\lstprompt{ssi >}$ verbose writel x x
$\lstprompt{ssi >}$ probe
Line 253: writel(val, pc->base + reg) => 0, 7e20004c
Line 253: writel(val, pc->base + reg) => 0, 7e200058
Line 253: writel(val, pc->base + reg) => 0, 7e200064
Line 253: writel(val, pc->base + reg) => 0, 7e200070
Line 253: writel(val, pc->base + reg) => 0, 7e20007c
Line 253: writel(val, pc->base + reg) => 0, 7e200088
Line 253: writel(val, pc->base + reg) => 0, 7e200050
Line 253: writel(val, pc->base + reg) => 0, 7e20005c
Line 253: writel(val, pc->base + reg) => 0, 7e200068
...
$\lstprompt{ssi >}$ b 498
$\lstprompt{ssi >}$ enable-irq 3
$\lstprompt{ssi :: On line 498}$
$\lstprompt{ssi >}$ xc offset
(1104, 0) = 3
$\lstprompt{ssi >}$ c
Line 253: writel(val, pc->base + reg) => 8, 7e20004c
$\lstprompt{ssi >}$
    \end{lstlisting}
    \caption{REPL interface to the driver's SSI}
    \label{fig:bcm2835-repl}
    \end{subfigure}
    \caption{Understanding modules in a large megasystem, such as the Linux
    device driver excerpted on the left hand side, is made more challenging by
    the difficulty of building, running, and tracing the code. A
    system-specific interpreter, such as our prototype demonstrated on the
    right hand side, allows one to directly execute individual modules in a
    simulated environment while tracing code behavior and memory values.}
    \label{fig:bcm2835}
\end{figure}

\subsection{Using a System-Specific Interpreter}
Suppose instead that Jane had a system-specific interpreter (SSI) for this
device driver, as shown in~\pref{fig:bcm2835-repl}. The SSI loads the driver
source code, then provides a \texttt{gdb}-like REPL interface to a simulated
instance of the driver.  Jane can ask it to load the driver and run different
driver operations in a simulated environment. She can trace, set breakpoints,
and step-through the driver code. \pref{fig:bcm2835-repl} shows how Jane can
use the SSI to print out the arguments of every call to \texttt{writel}, and
set a breakpoint to inspect the values of local variables.

Because the SSI is directly interpreting the driver code itself, not the entire
Linux kernel, it starts and executes quickly. If Jane later modifies the code,
she can immediately re-run the SSI to verify it produces the desired behavior.
Only once Jane has understood the driver code and made any desired
modifications might she need to worry about building and loading the driver.
The SSI allows her to separate concerns: understanding and modifying an
individual module in the megasystem without first understanding the entire
megasystem and associated build system.

\subsection{Writing a System-Specific Interpreter}
More likely, nobody has yet written an SSI for kernel drivers, but Jane has
heard of the concept and thinks it may help her task. She starts with a
pre-written \emph{SSI framework}, which has a parser for and implementation of
the base C language. Because the semantics of the underlying language are
handled by the SSI framework, Jane only needs to write the user-facing
interface (what should be run when the user requests to enable interrupts for a
pin?) and handlers for methods that call out to the larger megasystem API (what
should \texttt{of\_address\_to\_resource} return?).

\begin{figure}[t]
    \centering
    \begin{subfigure}[b]{0.48\linewidth}
        \begin{lstlisting}[linewidth=0.95\linewidth,basicstyle=\tiny]
def of_address_to_resource(np, which_resource, ptr_to_iomem):
  interpreter.exec_c(f"*{{0}} = (opaque)", ptr_to_iomem.cval())
  return interpreter.emit("(str (imm {0}))", 0)
        \end{lstlisting}
        \caption{Incomplete model of \texttt{of\_address\_to\_resource}}
        \label{fig:bad-model}
    \end{subfigure}
    \begin{subfigure}[b]{0.48\linewidth}
        \newcommand\lstprompt[1]{\color{white}{\textbf{#1}}}
        \begin{lstlisting}[backgroundcolor={\color{black!90}},basicstyle=\tiny\color{gray!30},morekeywords={ssi,>},numbers=none,mathescape=true]
...
$\lstprompt{ssi >}$ verbose writel x x
$\lstprompt{ssi >}$ probe
Line 253: Could not verbose because missing of_address_to_resource ( np , 0 , & iomem ) on line 1212
...
        \end{lstlisting}
        \caption{SSI warning when a model is incomplete}
        \label{fig:ssi-missing-model}
    \end{subfigure}
    \begin{subfigure}[b]{0.48\linewidth}
        \begin{lstlisting}[linewidth=0.95\linewidth,basicstyle=\tiny]
def of_address_to_resource(np, which_resource, ptr_to_iomem):
  addr = dtsi_find("example_pinctrl/bcm283x.dtsi", chosen_device)
  interpreter.exec_c(f"*{{0}} = {addr}", ptr_to_iomem.cval())
  return interpreter.emit("(str (imm {0}))", 0)
        \end{lstlisting}
        \caption{More complete model of the same method}
        \label{fig:good-model}
    \end{subfigure}
    \begin{subfigure}[b]{0.48\linewidth}
        \newcommand\lstprompt[1]{\color{white}{\textbf{#1}}}
        \begin{lstlisting}[backgroundcolor={\color{black!90}},basicstyle=\tiny\color{gray!30},morekeywords={ssi,>},numbers=none,mathescape=true]
...
$\lstprompt{ssi >}$ verbose writel x x
$\lstprompt{ssi >}$ probe
Line 253: writel(val, pc->base + reg) => 0, 7e20004c
...
        \end{lstlisting}
        \caption{SSI after improving the model}
        \label{fig:good-ssi}
    \end{subfigure}
    \caption{Users implement SSIs iteratively, refining their model of the
    system-module interface until the questions they want to ask can be
    answered. Jane's initial model of the \texttt{of\_address\_to\_resource(np,
    which\_resource, ptr\_to\_iomem)} sets \texttt{*ptr\_to\_iomem} to a fresh
    symbolic value and returns zero~(\pref{fig:bad-model}). But when she
    attempts to log memory writes during driver execution, the SSI informs her
    that the location and value of some of those writes depend on the value
    assigned to \texttt{*c} in the method~(\pref{fig:ssi-missing-model}). She
    refines the model~(\pref{fig:good-model}) until the SSI can provide the
    concrete information she is searching for~(\pref{fig:good-ssi}).}
\label{fig:interactive} \end{figure}

By default, the SSI framework will execute even in the presence of unknown
methods by tracking their return values as symbolic unknowns. If these values
do not affect the behavior Jane is interested in, she never needs to model
those methods. If they do affect some value Jane asks the SSI to print, the SSI
will helpfully point her towards exactly what missing API method was
missing~(\pref{fig:ssi-missing-model}).
This highlights an important feature of SSIs: the model of the surrounding
megasystem that Jane writes does not need to exactly match that of the original
system; it just needs to be \emph{operationally indistinguishable} with respect
to the questions that Jane wants to ask about this individual module.

In this way, Jane implements the SSI in an iterative, interactive
fashion~(\pref{fig:interactive}).  She starts off with the raw SSI framework,
which informs her when certain methods must be modeled to trace the behavior
she is concerned with. She then reads the code and documentation, forming a
mental model of those methods and mechanizing that model within the SSI. She
repeats this process, iteratively growing and mechanizing her mental model of
the system-module interface with the assistance of the SSI framework.

\section{Implementation of Prototype Interpreter Framework}
\label{sec:Framework}
We have implemented a prototype framework for building SSIs. The base framework
is roughly $1{,}000$ lines of Python code. Implementing the driver-specific SSI
on top of this framework shown in \pref{fig:bcm2835} takes approximately
another 100 lines of Python code. The first key idea is to separate
interpretation of the language-level syntax and semantics from the system-level
syntax and semantics.  This allows the SSI developer to focus on understanding
and implementing the system-level constructs without having to worry about the
core language-level constructs (such as addition, pointer dereference, etc.)
that are shared by all programs. The second key idea is to be as resilient to
missing semantics as possible. This allows the SSI developer to implement only
the minimum system-level semantics needed for the behavior they are interested
in.  System-level semantics that \emph{do not} affect that behavior should be
safely ignored by the interpreter. The rest of this section describes how our
prototype framework implements these key ideas.

\subsection{Parsing}
SSIs must be able to reasonably parse the code even if the user does not know
how to build it, e.g., we must handle missing files and headers. In this
context, C is known to be ambiguous to parse, e.g., the parsing of \texttt{T * x;}
depends on whether \texttt{T} is a typedef or a global variable. The same
issues arise when system-specific macros are used or when domain-specific
languages, such as DTSI, are compiled into the system during the build phase.
More broadly, we would like the parser to be flexible enough to allow the
interpreter to assist even if the program is being edited and not currently in
a valid syntactic state.

Our two key parsing mottos are: \emph{parse only what you need} and \emph{parse
compositionally}. We start parsing exactly where the user asked us to run. We
never parse code that is irrelevant to what the user has asked us to run. When
parsing, we do not use a single, monolithic grammar. Instead, we have a
hierarchy of rules that can match, e.g., an \texttt{if} statement rule or a
\texttt{while} statement rule. When these rules match, they (almost) never
recurse into other rules. Instead, like island grammars and
microgrammars~\citep{islandgrammars,microgrammars}, we parse with holes. A
simplified \texttt{if} statement rule looks for the keyword \texttt{if},
followed by a balanced set of parenthesis, followed by a balanced set of curly
brackets. The contents of the parentheses and brackets need not be valid C
code, just balanced. Then, upon executing the \texttt{if} statement, we
re-execute the parser on the contents of the condition and bracketed code. By
keeping the parsers extremely compositional in this form, the user can insert
their own custom parsing rules into the grammar hierarchy without fear of
messing up assumptions made by a monolithic, integrated grammar. Similarly, by
parsing compositionally with holes, if we do not end up taking the positive
branch of the \texttt{if} statement, we completely avoid having to deal with
any syntax errors within the branch body.

\subsection{Interpretation}
\label{sec:Tool_Interpretation}
Whereas in the parsing stage we needed to handle missing, invalid, or
system-specific syntactic constructs, in the interpretation stage we need to
handle missing, invalid, or system-specific \emph{semantic} constructs.

The simplest approach we provide to address this is to allow the programmer to
hook in to the interpretation phase to provide system-specific semantics. For
example, the user can write an interpreter-level function that is called
whenever \texttt{of\_address\_to\_resource} is encountered in the code, making
the correct changes to program state.

The second approach we provide is to simply \emph{ignore} certain values by
treating them as symbolic. By default, all memory values are symbolic.  As
operations are performed on these objects, new symbolic values are formed via
term-like compositions.  For example, executing \texttt{int x = a + b;}
would create new memory locations for $x$, $a$, $b$, with symbolic values $v_x,
v_a, v_b$, and a constraint $v_x = v_a + v_b$. If the concrete value of a
variable is identified, e.g., from a constant assignment or taking a branch
comparing against a constant, we propagate the constant greedily.  For example,
if we \emph{first} executed \texttt{int a = 1, b = 0;} then we would indeed
have set $v_x = 1$.

Why perform such a semi-symbolic execution, if the user ultimately cares about
the concrete values? The key insight is that the user usually only cares about
a \emph{subset} of the values, e.g., what values are written to which MMIO
memory locations.  Many values during a program execution do not affect those
that the user is interested in. For example, the \texttt{bcm2835} driver
interlaces spinlock, memory-management, and error-logging operations with the
main code, but these do not affect the actual values written to the
GPIO-control MMIO addresses. Leaving such values symbolic allows the user to
implement only those system-module interface methods that \emph{actually}
impact the subset of behavior they care about.

Because the code is interpreted, a number of interesting analyses can be
quickly implemented on top of an SSI. For example, every time a new symbolic or
concrete value is created during execution, the SSI tracks the values it was
created from along with the line being executed when it was created. At any
point, the user can request the interpreter produce a trace explaining the
program steps that led to this value. Such analyses are relatively
straight-forward to implement in SSIs built with our framework and can
significantly improve the debugging experience.




\section{Related Work}
SSIs bring megasystem-editing in languages like C closer to the live
programming environment dream~\citep{live_programming,vivalang}. They empower
the programmer to run incomplete code and evaluate the effect of modifications
much more quickly than a full build-run-debug cycle.  Differences with
traditional live programming include highlighting the use of
system-specific information; presenting a \texttt{gdb}-style interface
rather than full editor; and focusing on the benefits that programmers get by
simply writing such an SSI in the first place, even before using it.

SSIs are also similar to unit tests that mock methods external to the
module~\citep{unit_testing}.
SSIs essentially allow the unit tests to be written entirely separate from the
code and do not require the user wait for a full system build. Further, because
SSIs are generally written in a higher-level language, mocks can be written
using higher-level language features. The interpreter approach also allows the
user to more easily inspect and modify program execution.

The Alloy~\citep{alloy} and Redex~\citep{redex} projects show the power of
lightweight and partial mechanization of mental models of software for
understanding and debugging those models.
Running parts of existing code in isolation has been studied in the context of
\emph{microexecutions} for binary testing~\citep{microexec} and
\emph{under-constrained symbolic execution} for kernel bug
finding~\citep{ucklee}, inspiring the semi-symbolic interpretation
in~\pref{sec:Tool_Interpretation}.

Language server protocols allow certain questions about software to be
answered, such as the type of a variable or what methods may be called at any
given point in the program~\citep{lsp}.
%

C interpreters~\citep{vasilev2012cling,ccons} and frameworks for building
interpreters~\cite{interpframework} exist but, to our knowledge, are not
designed to incorporate the sort of system-specific information and flexible,
failure-tolerant interpretation described in~\pref{sec:Framework}.

The ``systems-as-languages'' way of thinking about systems is by now folklore
in the software community~\citep{paip,hints} and embraced by the
language-oriented programming language Racket~\citep{racket}. We aim to bring
this paradigm to megasystems written in imperative languages like C. Given the
ubiquitousness of this way of thinking about systems as domain-specific
languages, it is surprising that no prior work appears to have attempted to
directly \emph{interpret} that implicitly-defined DSL. The closest work we are
aware of is a line of research incorporating system-specific information into
bug-checkers and compiler
optimizations~\citep{mc_pldi,mc_osdi,interface_compilation,app_semantics}.

Alternative approaches to analyzing incomplete code either apply neural
networks~\citep{deep_parse} or attempt to \emph{repair} the code, automatically
inserting a minimal amount of new \texttt{typedef}s and method stubs to allow
the code to build with a standard
compiler~\citep{partial_java,inferring_static}. In theory, this avoids the need
of the SSI framework to re-implement the base language semantics. We initially
attempted to use the tool from~\citep{inferring_static} on the C driver code,
but found that it was unable to correct the errors, especially when macros were
used. We encountered similar difficulties with our own attempts at
implementing such an approach.  This experience motivated the flexible
parsing and interpretation approach described in~\pref{sec:Framework}, where
the interpreter can simply ignore any parts of the code it does not actually
attempt to run.

\section{Conclusion, Limitations, and Future Work}
The complexity of modern megasystems poses challenges to user freedom and
control over their computing. We propose system-specific interpreters (SSIs) to
address some of these challenges. SSIs provide users with more immediate
feedback than a full build-run-trace cycle, helping them better understand and
modify megasystem source code. Our prototype framework simplifies the
implementation of SSIs by decoupling the syntax and semantics of the language
and system.

SSIs are not perfect. Someone must write an SSI for the system, and keep it
up-to-date as the system evolves. Base languages with more complicated
semantics, such as C++ and Rust, will be more difficult to support. Although
semi-symbolic execution helps mitigate some of these issues, it introduces new
challenges like deciding on control flow when branching on symbolic values.


Once SSIs are available, new use cases for them may be found.  Direct
integration with IDEs may simplify the user experience.  Similarly, SSIs could
be used as a semantic code search tool, or to evaluate reasonableness of
suggestions from autocompletion tools in the context of this specific system.
Finally, they may be used by to extract preconditions for code, enabling better
static analysis and profile-guided optimizations.

\begin{acks}
    The author is grateful for financial support from the National Science
    Foundation under grants 1918056 and DGE-1656518. The author would also like
    to thank the anonymous reviewers for feedback that improved the quality of
    this paper. This paper benefited greatly from separate discussions the
    author had with Aditya V.\ Thakur and Zachary Yedidia regarding the problems
    posed by megasystems, as well as Dawson Engler regarding system-specific
    analyses.

    Icons in~\pref{fig:flowchart} were designed by OpenMoji, the open-source
    emoji and icon project, and licensed under CC BY-SA $4{.}0$. The OpenMoji
    project is available on the Web at \url{https://openmoji.org/} and the
    licence can be read at
    \url{https://creativecommons.org/licenses/by-sa/4.0/}.
\end{acks}

\bibliography{main}


\begin{thebibliography}{23}


\ifx \showCODEN    \undefined \def \showCODEN     #1{\unskip}     \fi
\ifx \showDOI      \undefined \def \showDOI       #1{#1}\fi
\ifx \showISBNx    \undefined \def \showISBNx     #1{\unskip}     \fi
\ifx \showISBNxiii \undefined \def \showISBNxiii  #1{\unskip}     \fi
\ifx \showISSN     \undefined \def \showISSN      #1{\unskip}     \fi
\ifx \showLCCN     \undefined \def \showLCCN      #1{\unskip}     \fi
\ifx \shownote     \undefined \def \shownote      #1{#1}          \fi
\ifx \showarticletitle \undefined \def \showarticletitle #1{#1}   \fi
\ifx \showURL      \undefined \def \showURL       {\relax}        \fi
\providecommand\bibfield[2]{#2}
\providecommand\bibinfo[2]{#2}
\providecommand\natexlab[1]{#1}
\providecommand\showeprint[2][]{arXiv:#2}

\bibitem[Ahmed et~al\mbox{.}(2021)]%
        {deep_parse}
\bibfield{author}{\bibinfo{person}{Toufique Ahmed},
  \bibinfo{person}{Premkumar~T. Devanbu}, {and} \bibinfo{person}{Vincent~J.
  Hellendoorn}.} \bibinfo{year}{2021}\natexlab{}.
\newblock \showarticletitle{Learning lenient parsing {\&} typing via indirect
  supervision}.
\newblock \bibinfo{journal}{\emph{Empir. Softw. Eng.}} \bibinfo{volume}{26},
  \bibinfo{number}{2} (\bibinfo{year}{2021}), \bibinfo{pages}{29}.
\newblock
\urldef\tempurl%
\url{https://doi.org/10.1007/s10664-021-09942-y}
\showDOI{\tempurl}


\bibitem[Brown et~al\mbox{.}(2020)]%
        {microgrammars}
\bibfield{author}{\bibinfo{person}{Fraser Brown}, \bibinfo{person}{Deian
  Stefan}, {and} \bibinfo{person}{Dawson~R. Engler}.}
  \bibinfo{year}{2020}\natexlab{}.
\newblock \showarticletitle{Sys: {A} Static/Symbolic Tool for Finding Good Bugs
  in Good (Browser) Code}. In \bibinfo{booktitle}{\emph{29th {USENIX} Security
  Symposium, {USENIX} Security 2020, August 12-14, 2020}},
  \bibfield{editor}{\bibinfo{person}{Srdjan Capkun} {and}
  \bibinfo{person}{Franziska Roesner}} (Eds.). \bibinfo{publisher}{{USENIX}
  Association}, \bibinfo{pages}{199--216}.
\newblock
\urldef\tempurl%
\url{https://www.usenix.org/conference/usenixsecurity20/presentation/brown}
\showURL{%
\tempurl}


\bibitem[Dagenais and Hendren(2008)]%
        {partial_java}
\bibfield{author}{\bibinfo{person}{Barth{\'{e}}l{\'{e}}my Dagenais} {and}
  \bibinfo{person}{Laurie~J. Hendren}.} \bibinfo{year}{2008}\natexlab{}.
\newblock \showarticletitle{Enabling static analysis for partial java
  programs}. In \bibinfo{booktitle}{\emph{Proceedings of the 23rd Annual {ACM}
  {SIGPLAN} Conference on Object-Oriented Programming, Systems, Languages, and
  Applications, {OOPSLA} 2008, October 19-23, 2008, Nashville, TN, {USA}}},
  \bibfield{editor}{\bibinfo{person}{Gail~E. Harris}} (Ed.).
  \bibinfo{publisher}{{ACM}}, \bibinfo{pages}{313--328}.
\newblock
\urldef\tempurl%
\url{https://doi.org/10.1145/1449764.1449790}
\showDOI{\tempurl}


\bibitem[Daka and Fraser(2014)]%
        {unit_testing}
\bibfield{author}{\bibinfo{person}{Ermira Daka} {and} \bibinfo{person}{Gordon
  Fraser}.} \bibinfo{year}{2014}\natexlab{}.
\newblock \showarticletitle{A Survey on Unit Testing Practices and Problems}.
  In \bibinfo{booktitle}{\emph{25th {IEEE} International Symposium on Software
  Reliability Engineering, {ISSRE} 2014, Naples, Italy, November 3-6, 2014}}.
  \bibinfo{publisher}{{IEEE} Computer Society}, \bibinfo{pages}{201--211}.
\newblock
\urldef\tempurl%
\url{https://doi.org/10.1109/ISSRE.2014.11}
\showDOI{\tempurl}


\bibitem[Durham et~al\mbox{.}(2003)]%
        {interpframework}
\bibfield{author}{\bibinfo{person}{Alan~M. Durham}, \bibinfo{person}{Edson
  Sussumu}, {and} \bibinfo{person}{Arlindo~Fl{\'{a}}vio da
  Concei{\c{c}}{\~{a}}o}.} \bibinfo{year}{2003}\natexlab{}.
\newblock \showarticletitle{A framework for building language interpreters}. In
  \bibinfo{booktitle}{\emph{Companion of the 18th Annual {ACM} {SIGPLAN}
  Conference on Object-Oriented Programming, Systems, Languages, and
  Applications, {OOPSLA} 2003, October 26-30, 2003, Anaheim, CA, {USA}}},
  \bibfield{editor}{\bibinfo{person}{Ron Crocker} {and} \bibinfo{person}{Guy
  L.~Steele Jr.}} (Eds.). \bibinfo{publisher}{{ACM}},
  \bibinfo{pages}{191--196}.
\newblock
\urldef\tempurl%
\url{https://doi.org/10.1145/949344.949398}
\showDOI{\tempurl}


\bibitem[Engler(1997)]%
        {app_semantics}
\bibfield{author}{\bibinfo{person}{Dawson~R. Engler}.}
  \bibinfo{year}{1997}\natexlab{}.
\newblock \showarticletitle{Incorporating Application Semantics and Control
  into Compilation}. In \bibinfo{booktitle}{\emph{Proceedings of the Conference
  on Domain-Specific Languages, DSL'97, Santa Barbara, California, USA, October
  15-17, 1997}}, \bibfield{editor}{\bibinfo{person}{Chris Ramming}} (Ed.).
  \bibinfo{publisher}{{USENIX}}.
\newblock
\urldef\tempurl%
\url{http://www.usenix.org/publications/library/proceedings/dsl97/engler.html}
\showURL{%
\tempurl}


\bibitem[Engler(1999)]%
        {interface_compilation}
\bibfield{author}{\bibinfo{person}{Dawson~R. Engler}.}
  \bibinfo{year}{1999}\natexlab{}.
\newblock \showarticletitle{Interface Compilation: Steps Toward Compiling
  Program Interfaces as Languages}.
\newblock \bibinfo{journal}{\emph{{IEEE} Trans. Software Eng.}}
  \bibinfo{volume}{25}, \bibinfo{number}{3} (\bibinfo{year}{1999}),
  \bibinfo{pages}{387--400}.
\newblock
\urldef\tempurl%
\url{https://doi.org/10.1109/32.798327}
\showDOI{\tempurl}


\bibitem[Engler et~al\mbox{.}(2000)]%
        {mc_osdi}
\bibfield{author}{\bibinfo{person}{Dawson~R. Engler}, \bibinfo{person}{Benjamin
  Chelf}, \bibinfo{person}{Andy Chou}, {and} \bibinfo{person}{Seth Hallem}.}
  \bibinfo{year}{2000}\natexlab{}.
\newblock \showarticletitle{Checking System Rules Using System-Specific,
  Programmer-Written Compiler Extensions}. In \bibinfo{booktitle}{\emph{4th
  Symposium on Operating System Design and Implementation {(OSDI} 2000), San
  Diego, California, USA, October 23-25, 2000}},
  \bibfield{editor}{\bibinfo{person}{Michael~B. Jones} {and}
  \bibinfo{person}{M.~Frans Kaashoek}} (Eds.). \bibinfo{publisher}{{USENIX}
  Association}, \bibinfo{pages}{1--16}.
\newblock
\urldef\tempurl%
\url{http://dl.acm.org/citation.cfm?id=1251230}
\showURL{%
\tempurl}


\bibitem[Felleisen et~al\mbox{.}(2015)]%
        {racket}
\bibfield{author}{\bibinfo{person}{Matthias Felleisen},
  \bibinfo{person}{Robert~Bruce Findler}, \bibinfo{person}{Matthew Flatt},
  \bibinfo{person}{Shriram Krishnamurthi}, \bibinfo{person}{Eli Barzilay},
  \bibinfo{person}{Jay~A. McCarthy}, {and} \bibinfo{person}{Sam
  Tobin{-}Hochstadt}.} \bibinfo{year}{2015}\natexlab{}.
\newblock \showarticletitle{The Racket Manifesto}. In
  \bibinfo{booktitle}{\emph{1st Summit on Advances in Programming Languages,
  {SNAPL} 2015, May 3-6, 2015, Asilomar, California, {USA}}}
  \emph{(\bibinfo{series}{LIPIcs}, Vol.~\bibinfo{volume}{32})},
  \bibfield{editor}{\bibinfo{person}{Thomas Ball}, \bibinfo{person}{Rastislav
  Bod{\'{\i}}k}, \bibinfo{person}{Shriram Krishnamurthi},
  \bibinfo{person}{Benjamin~S. Lerner}, {and} \bibinfo{person}{Greg Morrisett}}
  (Eds.). \bibinfo{publisher}{Schloss Dagstuhl - Leibniz-Zentrum f{\"{u}}r
  Informatik}, \bibinfo{pages}{113--128}.
\newblock
\urldef\tempurl%
\url{https://doi.org/10.4230/LIPIcs.SNAPL.2015.113}
\showDOI{\tempurl}


\bibitem[Godefroid(2014)]%
        {microexec}
\bibfield{author}{\bibinfo{person}{Patrice Godefroid}.}
  \bibinfo{year}{2014}\natexlab{}.
\newblock \showarticletitle{Micro execution}. In \bibinfo{booktitle}{\emph{36th
  International Conference on Software Engineering, {ICSE} '14, Hyderabad,
  India - May 31 - June 07, 2014}}, \bibfield{editor}{\bibinfo{person}{Pankaj
  Jalote}, \bibinfo{person}{Lionel~C. Briand}, {and}
  \bibinfo{person}{Andr{\'{e}} van~der Hoek}} (Eds.).
  \bibinfo{publisher}{{ACM}}, \bibinfo{pages}{539--549}.
\newblock
\urldef\tempurl%
\url{https://doi.org/10.1145/2568225.2568273}
\showDOI{\tempurl}


\bibitem[Gunasinghe and Marcus(2022)]%
        {lsp}
\bibfield{author}{\bibinfo{person}{Nadeeshaan Gunasinghe} {and}
  \bibinfo{person}{Nipuna Marcus}.} \bibinfo{year}{2022}\natexlab{}.
\newblock \bibinfo{title}{Language Server Protocol and Implementation}.
\newblock
\newblock


\bibitem[Hallem et~al\mbox{.}(2002)]%
        {mc_pldi}
\bibfield{author}{\bibinfo{person}{Seth Hallem}, \bibinfo{person}{Benjamin
  Chelf}, \bibinfo{person}{Yichen Xie}, {and} \bibinfo{person}{Dawson~R.
  Engler}.} \bibinfo{year}{2002}\natexlab{}.
\newblock \showarticletitle{A System and Language for Building System-Specific,
  Static Analyses}. In \bibinfo{booktitle}{\emph{Proceedings of the 2002 {ACM}
  {SIGPLAN} Conference on Programming Language Design and Implementation
  (PLDI), Berlin, Germany, June 17-19, 2002}},
  \bibfield{editor}{\bibinfo{person}{Jens Knoop} {and}
  \bibinfo{person}{Laurie~J. Hendren}} (Eds.). \bibinfo{publisher}{{ACM}},
  \bibinfo{pages}{69--82}.
\newblock
\urldef\tempurl%
\url{https://doi.org/10.1145/512529.512539}
\showDOI{\tempurl}


\bibitem[Jackson(2019)]%
        {alloy}
\bibfield{author}{\bibinfo{person}{Daniel Jackson}.}
  \bibinfo{year}{2019}\natexlab{}.
\newblock \showarticletitle{Alloy: a Language and Tool for Exploring Software
  Designs}.
\newblock \bibinfo{journal}{\emph{{Communications} of the {ACM}}}
  \bibinfo{volume}{62}, \bibinfo{number}{9} (\bibinfo{year}{2019}),
  \bibinfo{pages}{66--76}.
\newblock
\urldef\tempurl%
\url{https://doi.org/10.1145/3338843}
\showDOI{\tempurl}


\bibitem[Klein et~al\mbox{.}(2012)]%
        {redex}
\bibfield{author}{\bibinfo{person}{Casey Klein}, \bibinfo{person}{John
  Clements}, \bibinfo{person}{Christos Dimoulas}, \bibinfo{person}{Carl
  Eastlund}, \bibinfo{person}{Matthias Felleisen}, \bibinfo{person}{Matthew
  Flatt}, \bibinfo{person}{Jay~A. McCarthy}, \bibinfo{person}{Jon Rafkind},
  \bibinfo{person}{Sam Tobin{-}Hochstadt}, {and} \bibinfo{person}{Robert~Bruce
  Findler}.} \bibinfo{year}{2012}\natexlab{}.
\newblock \showarticletitle{Run your research: on the effectiveness of
  lightweight mechanization}. In \bibinfo{booktitle}{\emph{Proceedings of the
  39th {ACM} {SIGPLAN-SIGACT} Symposium on Principles of Programming Languages,
  {POPL} 2012}}, \bibfield{editor}{\bibinfo{person}{John Field} {and}
  \bibinfo{person}{Michael Hicks}} (Eds.).
\newblock
\urldef\tempurl%
\url{https://doi.org/10.1145/2103656.2103691}
\showDOI{\tempurl}


\bibitem[Lampson(1983)]%
        {hints}
\bibfield{author}{\bibinfo{person}{Butler~W. Lampson}.}
  \bibinfo{year}{1983}\natexlab{}.
\newblock \showarticletitle{Hints for Computer System Design}. In
  \bibinfo{booktitle}{\emph{Proceedings of the Ninth {ACM} Symposium on
  Operating System Principles, {SOSP} 1983, Bretton Woods, New Hampshire, USA,
  October 10-13, 1983}}, \bibfield{editor}{\bibinfo{person}{Jerome~H. Saltzer},
  \bibinfo{person}{Roy Levin}, {and} \bibinfo{person}{David~D. Redell}} (Eds.).
  \bibinfo{publisher}{{ACM}}, \bibinfo{pages}{33--48}.
\newblock
\urldef\tempurl%
\url{https://doi.org/10.1145/800217.806614}
\showDOI{\tempurl}


\bibitem[Melo et~al\mbox{.}(2018)]%
        {inferring_static}
\bibfield{author}{\bibinfo{person}{Leandro T.~C. Melo},
  \bibinfo{person}{Rodrigo~Geraldo Ribeiro}, \bibinfo{person}{Marcus~R. de
  Ara{\'{u}}jo}, {and} \bibinfo{person}{Fernando Magno~Quint{\~{a}}o Pereira}.}
  \bibinfo{year}{2018}\natexlab{}.
\newblock \showarticletitle{Inference of static semantics for incomplete {C}
  programs}.
\newblock \bibinfo{journal}{\emph{Proc. {ACM} Program. Lang.}}
  \bibinfo{volume}{2}, \bibinfo{number}{{POPL}} (\bibinfo{year}{2018}),
  \bibinfo{pages}{29:1--29:28}.
\newblock
\urldef\tempurl%
\url{https://doi.org/10.1145/3158117}
\showDOI{\tempurl}


\bibitem[Nilsson{-}Nyman et~al\mbox{.}(2008)]%
        {islandgrammars}
\bibfield{author}{\bibinfo{person}{Emma Nilsson{-}Nyman},
  \bibinfo{person}{Torbj{\"{o}}rn Ekman}, {and} \bibinfo{person}{G{\"{o}}rel
  Hedin}.} \bibinfo{year}{2008}\natexlab{}.
\newblock \showarticletitle{Practical Scope Recovery Using Bridge Parsing}. In
  \bibinfo{booktitle}{\emph{Software Language Engineering, First International
  Conference, {SLE} 2008, Toulouse, France, September 29-30, 2008. Revised
  Selected Papers}} \emph{(\bibinfo{series}{Lecture Notes in Computer Science},
  Vol.~\bibinfo{volume}{5452})}, \bibfield{editor}{\bibinfo{person}{Dragan
  Gasevic}, \bibinfo{person}{Ralf L{\"{a}}mmel}, {and}
  \bibinfo{person}{Eric~Van Wyk}} (Eds.). \bibinfo{publisher}{Springer},
  \bibinfo{pages}{95--113}.
\newblock
\urldef\tempurl%
\url{https://doi.org/10.1007/978-3-642-00434-6\_7}
\showDOI{\tempurl}


\bibitem[Norvig(1992)]%
        {paip}
\bibfield{author}{\bibinfo{person}{Peter Norvig}.}
  \bibinfo{year}{1992}\natexlab{}.
\newblock \bibinfo{booktitle}{\emph{Paradigms of artificial intelligence
  programming: case studies in Common LISP}}.
\newblock \bibinfo{publisher}{Morgan Kaufmann}.
\newblock


\bibitem[Ramos and Engler(2015)]%
        {ucklee}
\bibfield{author}{\bibinfo{person}{David~A. Ramos} {and}
  \bibinfo{person}{Dawson~R. Engler}.} \bibinfo{year}{2015}\natexlab{}.
\newblock \showarticletitle{Under-Constrained Symbolic Execution: Correctness
  Checking for Real Code}. In \bibinfo{booktitle}{\emph{24th {USENIX} Security
  Symposium, {USENIX} Security 15, Washington, D.C., USA, August 12-14, 2015}},
  \bibfield{editor}{\bibinfo{person}{Jaeyeon Jung} {and}
  \bibinfo{person}{Thorsten Holz}} (Eds.). \bibinfo{publisher}{{USENIX}
  Association}, \bibinfo{pages}{49--64}.
\newblock
\urldef\tempurl%
\url{https://www.usenix.org/conference/usenixsecurity15/technical-sessions/presentation/ramos}
\showURL{%
\tempurl}


\bibitem[Svitkine(2009)]%
        {ccons}
\bibfield{author}{\bibinfo{person}{Alexei Svitkine}.}
  \bibinfo{year}{2009}\natexlab{}.
\newblock \bibinfo{title}{Interactive Console for the C Programming Language}.
\newblock
\newblock


\bibitem[Tanimoto(1990)]%
        {vivalang}
\bibfield{author}{\bibinfo{person}{Steven~L. Tanimoto}.}
  \bibinfo{year}{1990}\natexlab{}.
\newblock \showarticletitle{{VIVA:} {A} visual language for image processing}.
\newblock \bibinfo{journal}{\emph{J. Vis. Lang. Comput.}} \bibinfo{volume}{1},
  \bibinfo{number}{2} (\bibinfo{year}{1990}), \bibinfo{pages}{127--139}.
\newblock
\urldef\tempurl%
\url{https://doi.org/10.1016/S1045-926X(05)80012-6}
\showDOI{\tempurl}


\bibitem[Tanimoto(2013)]%
        {live_programming}
\bibfield{author}{\bibinfo{person}{Steven~L. Tanimoto}.}
  \bibinfo{year}{2013}\natexlab{}.
\newblock \showarticletitle{A perspective on the evolution of live
  programming}. In \bibinfo{booktitle}{\emph{Proceedings of the 1st
  International Workshop on Live Programming, {LIVE} 2013, San Francisco,
  California, USA, May 19, 2013}}, \bibfield{editor}{\bibinfo{person}{Brian
  Burg}, \bibinfo{person}{Adrian Kuhn}, {and} \bibinfo{person}{Chris Parnin}}
  (Eds.). \bibinfo{publisher}{{IEEE} Computer Society},
  \bibinfo{pages}{31--34}.
\newblock
\urldef\tempurl%
\url{https://doi.org/10.1109/LIVE.2013.6617346}
\showDOI{\tempurl}


\bibitem[Vasilev et~al\mbox{.}(2012)]%
        {vasilev2012cling}
\bibfield{author}{\bibinfo{person}{Vasil Vasilev}, \bibinfo{person}{Ph Canal},
  \bibinfo{person}{Axel Naumann}, {and} \bibinfo{person}{Paul Russo}.}
  \bibinfo{year}{2012}\natexlab{}.
\newblock \showarticletitle{Cling--the new interactive interpreter for ROOT 6}.
  In \bibinfo{booktitle}{\emph{Journal of Physics: Conference Series}},
  Vol.~\bibinfo{volume}{396}. IOP Publishing, \bibinfo{pages}{052071}.
\newblock


\end{thebibliography}

%

\end{document}